# Electrically Interconnected Platinum Nanonetworks for Flexible Electronics

Sherjeel Mahmood Baig*, Hideki Abe*

**ABSTRACT:** Flexible electronics are attracting attention due to the growing demand for lightweight, bendable devices that can conform to various surfaces including human skin. Although indium tin oxide (ITO) is widely used for electrical interconnection in flexible electronics, its brittleness limits its durability under repeated bending. Here, we introduce platinum (Pt) nanonetworks as an alternative to ITO, offering superior electrical stability under intense and repeated bending conditions. Electrically interconnected Pt nanonetworks with an average thickness below 50 nm are fabricated on polyimide (PI) substrates via an atmospheric treatment that promotes nanophase separation in thin deposition films of a platinum-cerium (Pt-Ce) alloy, developing a nanotexture of Pt and insulating cerium dioxide ($CeO_2$). The resulting Pt nanonetworks on PI exhibit high mechanical flexibility, maintaining a sheet resistance of approximately 2.76 kΩ/sq even after 1000 bending cycles at varying diameters, down to 1.5 mm. Detailed characterization reveals critical temperature and time thresholds in the atmospheric treatment necessary to form interconnected Pt nanonetworks on solid surfaces: interconnected nanonetworks develop at lower temperatures and shorter treatment times, while higher temperatures and longer treatments lead to disconnected Pt nanoislands. LCR (Inductance, Capacitance, and Resistance) measurements further show that the interconnected Pt nanonetworks exhibit inductor-like electrical responses, while disconnected Pt nanoislands display capacitor-like behavior.

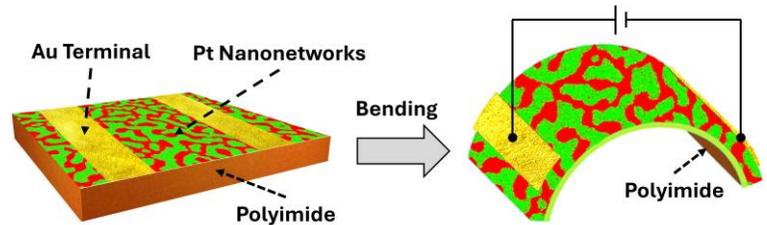

## INTRODUCTION

Flexible electronics have become a critical field due to the growing demand for lightweight, bendable, and stretchable devices that can be seamlessly integrated with various surfaces including human skin. From wearable health monitoring systems to foldable displays and touchscreens, flexible electronics require electroconductive materials capable of maintaining reliable performance even under significant mechanical strain. Traditional rigid materials, such as silicon (Si), are unsuitable for these applications due to their limited flexibility and vulnerability to mechanical failure.[1] Indium tin oxide (ITO), commonly used for electric interconnections in flexible dis-plays, is brittle,[2] prone to cracking,[3] And faces challenges due to the scarcity of indium, which hinders large-scale development.[4,5] Dong et al. utilized laser interference lithography (LIL) to create ITO nanopatterns that enable multiaxial bending while maintaining low electric resistance.[6] However, LIL is a complex, time-consuming, and low-throughput method.

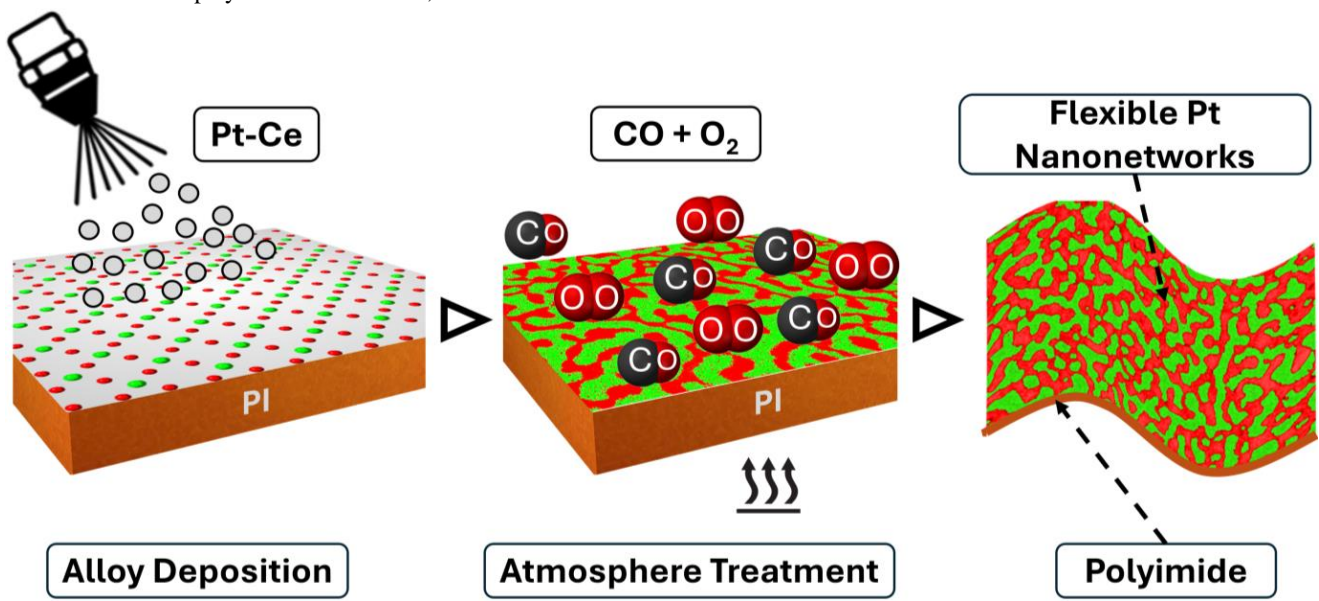

Figure 1. Schematic image for a device incorporating flexible platinum nanonetworks. Pt-Ce alloy was deposited over the Polyimide surface. Followed by the atmosphere treatment in the presence of CO and $O_2$ at the elevated temperature. The Pt nanonetworks (red) emerged via the oxidation of Ce into $CeO_2$ (green) over the PI substrate.

Metal nanomesh or nanonetworks have garnered increasing attention for their ability to combine mechanical flexibility with high electrical conductivity. Seo et al. demonstrated that gold (Au) nanomesh fabricated via nanosphere lithography outperformed ITO interconnections in electrophysiology applications, offering higher flexibility.[7] Guo et al. fabricated Au nanomesh on a flexible polydimethylsiloxane (PDMS) substrate using grain boundary lithography, a bilayer lift-off metallization process.[8] However, their methods remain complex and time-consuming, requiring multiple lithographic steps. Adrien and colleagues fabricated Au nanomesh on polyethylene terephthalate (PET) substrates via a chemical process, bypassing lithography, and showed good electrical conductivity and exceptional stability under mechanical deformation.[9] Unfortunately, their method has significant drawbacks, such as limitations in scalability due to the reliance on floating microdomains for nanomesh transfer and safety concerns related to the handling of concentrated nitric acid vapors during the dealloying process.

Here, we present a straightforward method for fabricating flexible platinum (Pt) nanonetworks not only on solid substrates but on flexible substrates as well (Fig. 1). The process begins with the deposition of 50 nm-thick platinum-cerium (Pt-Ce) alloy films onto the substrate surface, followed by an atmospheric treatment at elevated temperatures using a gas mixture of carbon monoxide (CO) and oxygen ($O_2$). This treatment drives phase separation in the Pt-Ce alloy, forming a nanotexture of Pt and cerium dioxide ($CeO_2$). The optimal conditions for producing highly interconnected Pt nanotextures were determined through a phase diagram created from atmospheric treatments applied to Pt-Ce films on silicon substrates, varying in temperature from 300–500 °C and treatment duration. Based on the phase diagram, electrically interconnected Pt nanonetworks were successfully fabricated on polyimide (PI) film substrates. These Pt nanonetworks on PI demonstrated outstanding electrical stability, maintaining a sheet resistance of 2.76 kΩ/sq even after 1000 bending cycles at a radius as small as 1.5 mm. The proposed fabrication method for Pt nanonetworks offers a practical approach to flexible electronics by using a simple protocol involving alloy film deposition and atmospheric treatments, enabling large-area electric interconnections without the need for complex processes or specialized equipment such as lithographs.

**RESULTS & DISCUSSION**

Figure 2 presents a phase diagram illustrating the nanotextures of Pt-Ce films on Si substrates, subjected to varied atmospheric treatments. FE-SEM images of these films are arranged by treatment temperature and duration, showing the impact of these conditions on nanotexture morphology. The films were exposed to a CO, $O_2$, and Ar gas mixture in a mole ratio of 2:1:97, with both temperature and treatment duration varying. An interconnected Pt nanotexture developed within the triangular region defined by the line from a treatment temperature of 400 °C and a duration of 60 minutes to the origin. The highest degree of interconnection appeared at 300 °C with a 30-minute duration, whereas no pattern was observed at 300 °C with only a 1-minute treatment. This suggests that temperatures below 300 °C or shorter durations lack the thermal energy needed to initiate the phase separation of Pt-Ce alloys required for forming the Pt and $CeO_2$ nanotexture.

The Pt and $CeO_2$ nanotexture underwent a morphological transition from interconnected Pt networks to isolated, discrete Pt islands as the treatment temperature exceeded 450 °C. This transition from networks to islands is attributed to the accelerated oxidation of Ce in the Pt-Ce alloy to $CeO_2$ at elevated temperatures, which further promotes Pt atom agglomeration, disrupting the connectivity of the Pt nanonetworks. Extended treatment durations also led to the formation of Pt islands, indicating increased Pt agglomeration over time. In contrast, shorter treatment durations limited Pt atom diffusion and clustering at all temperatures, thereby preserving the enhanced connectivity within the Pt nanonetwork.

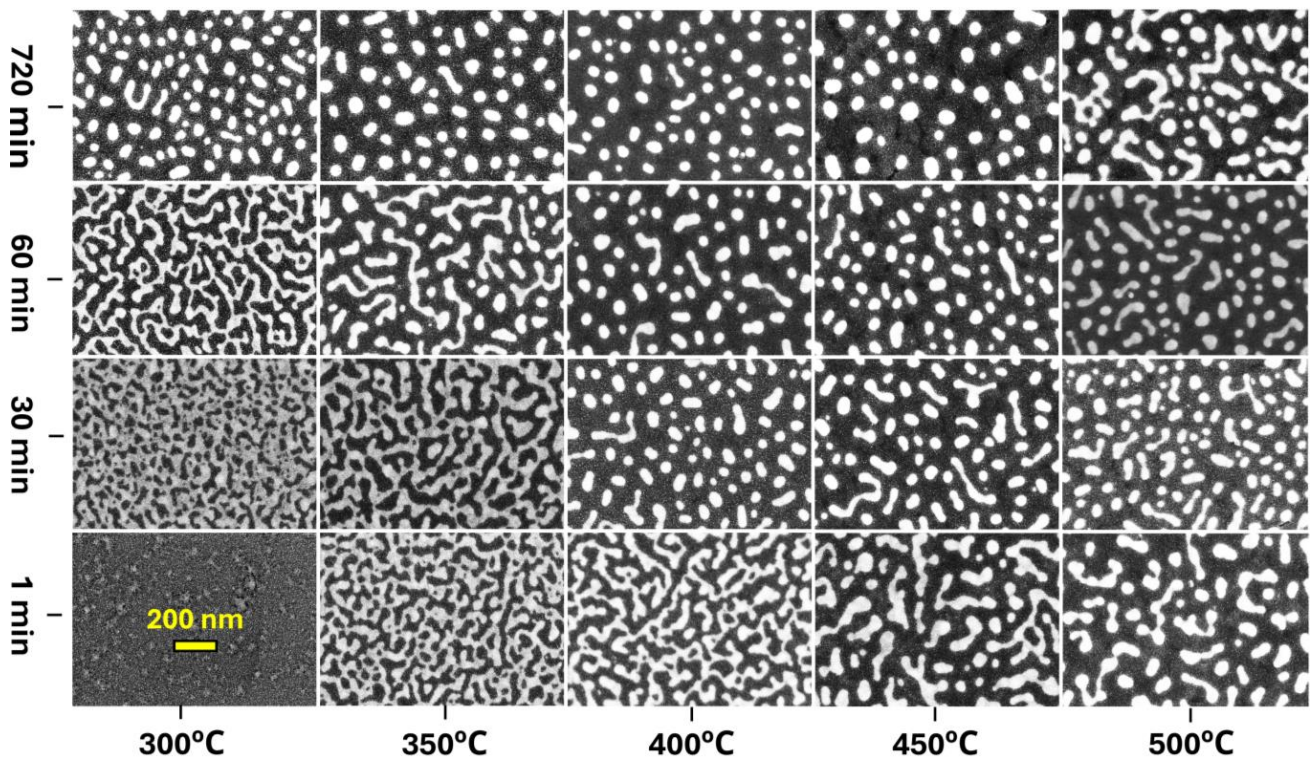

Figure 2. Phase diagram for atmospheric-treated Pt-Ce films, showing the morphological transition from interconnected networks to discrete nano-islands, as influenced by varying treatment temperatures (horizontal axis) and durations (vertical axis). The bright and dark areas correspond to the Pt- and $CeO_2$ phases, respectively.

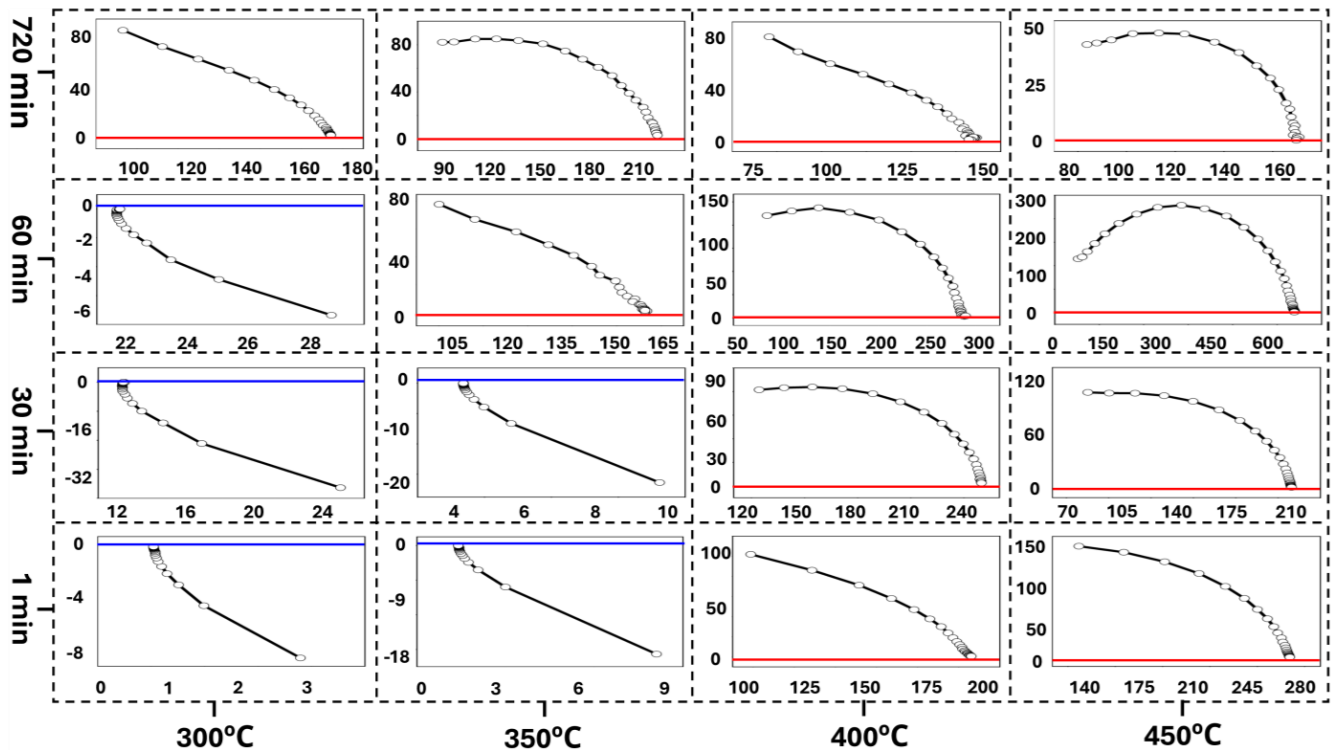

Figure 3. Nyquist plots for atmospheric-treated Pt-Ce films (see Figure 2), obtained from LCR measurements across a frequency range of 1–5 MHz. Cole-Cole arcs for interconnected nanotextures (blue baselines) are located in the lower half of the complex impedance plane, while arcs for island-like nanotextures (red baselines) appear in the upper half.

LCR measurements offered valuable insights into the relationship between electrical properties and the morphology of the Pt nanotextures. LCR measurements were performed over a frequency range of 1 Hz to 5 MHz. Each of the Nyquist plots in Figure 3 corresponds to the FE-SEM images in Figure 2. The interconnected Pt nanonetworks, observed at lower temperatures for short durations in Figure 2, exhibited inductor-like electrical behavior, as shown in Figure 3. The Cole-Cole arcs for these networks were positioned on the lower side of the complex impedance plane, typical for a parallel configuration of an inductor and resistor. By fitting the Cole-Cole arc with an equivalent circuit (Figure S1), the inductance and sheet resistance of the Pt nanonetworks were calculated to be 0.7 µH and 2.76 kΩ/sq, respectively.

The Pt nanoislands observed at temperatures between 300 °C and 450 °C for extended durations exhibited capacitor-like responses, in contrast to the inductor-like behavior of Pt nanonetworks. The Cole-Cole arcs for the Pt islands appeared on the upper half of the complex impedance plane, characteristic of a parallel circuit comprising a capacitor and resistor (Figure S2). The capacitance of the Pt islands was calculated to range from 0.2 to 0.6 nF, while the sheet resistance was 30 kΩ/sq. These Pt nanoislands, separated by insulating $CeO_2$, store charge between neighboring islands, but are unsuitable for applications involving direct current due to their high sheet resistance. The frequency-dependent response of each sample, along with the corresponding Z-fitted curves, is displayed in Figure S3. Impedance analysis up to 30MHz confirmed that inductive nanonetworks and capacitive nanoislands retained stability, even under extreme frequency testing (Figure S4 and S5).

Flexibility tests were conducted on Pt nanonetworks on polyimide films by subjecting them to repeated bending at various diameters (4.2 mm, 4.0 mm, 2.5 mm, 2.0 mm, and 1.5 mm), as shown in Figure S6. The bending setup and connection terminals are illustrated in Figure 4a, where conductive tape was used over the Au-Ti terminals to ensure smooth and non-destructive connections. Sheet resistance was measured at different bending diameters, revealing that the sheet resistance of the Pt nanonetworks on polyimide films remained approximately constant even after 1000 bending cycles at a bending radius as small as 1.5 mm. Minor resistance increase observed due to terminal cracking (see FE-SEM images in Fig. S7 and S8). This behavior is significantly better than that of traditional ITO thin films, which typically experience a rapid increase in resistance upon bending due to crack formation.

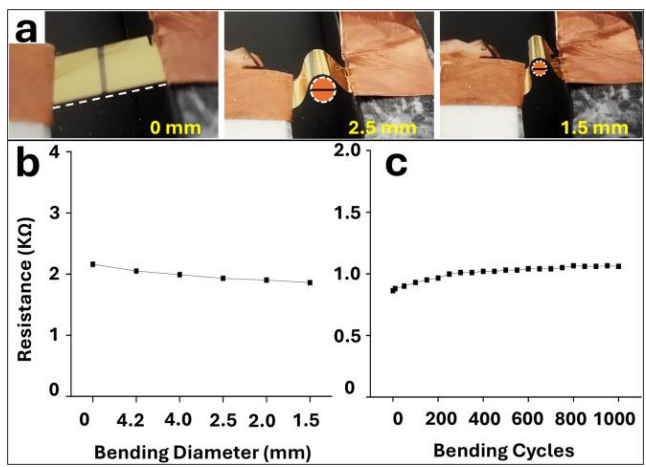

Figure 4. (a) Bending setup and connection terminals, with conductive tape contacting Au-Ti terminals. (b) Flexibility test showing the sheet resistance of the Pt nanonetwork sample at bending diameters down to 1.5 mm. (c) Cyclic bending test results for the Pt nanonetwork sample, demonstrating sheet resistance stability over 1000 cycles of bending at a diameter of 1.5 mm.

## CONCLUSIONS

We successfully demonstrated the fabrication of highly flexible and electrically interconnected Pt nanonetworks on solid substrates by a simple method using Pt-Ce alloy sputtering followed by atmospheric treatments. The Pt nanonetworks showed high electrical conductivity, mechanical flexibility, and excellent stability to repeated bending, serving as an ideal candidate material for flexible electronics. Furthermore, the proposed fabrication method is cost-effective, easy to implement, and does not require complex equipment, providing a promising approach for the scalable manufacturing of flexible electronic devices.

## METHODS

**Alloy synthesis and characterization**. A Pt-Ce alloy sputtering target was prepared by melting high-purity Pt (Furuya Kinzoku, 99.9 %) and Ce (Aldrich, 99.9 %) metal ingots in a mole ratio of Pt:Ce=2:1 using an arc torch in a pure Ar (99.9999 %) atmosphere. The phase purity of the resulting $Pt_2Ce$ alloy was confirmed via powder X-ray diffraction ($p$XRD, X'Pert Pro, Panalytical).[10]

**Alloy thin film deposition on Si and post-treatment:** Thin films of the Pt-Ce alloy, 50 ± 1 nm thick, were deposited onto Si substrates (380 μm thick Si wafer) at room temperature (see FE-SEM and EDX images in Fig. S9) using an electron-beam evaporator (MB-501010) (Fig. S10). Atmospheric treatments were applied to the Pt-Ce films at temperatures ranging from 300 °C to 500 °C for durations of 1 minute to 12 hours, in a controlled $O_2$ and CO atmosphere balanced with Ar at a volumetric ratio of 1:2:97 and a flow rate of 10 ml/min. The Pt-Ce films were converted into composite films of Pt and $CeO_2$ through atmospheric treatments, forming different nanotextures depending on the treatment conditions.

**Thin film characterization:** The nanotexture of the atmospheric-treated Pt-Ce films was analyzed to identify the optimal conditions for highly interconnected Pt nanonetworks. The analysis involved Grazing Incidence X-ray Diffraction (GIXRD) (Fig. S11), Field Emission Scanning Electron Microscopy (FE-SEM), and Energy-Dispersive X-ray (EDX) analysis (Fig. S12, and Atomic Force Microscopy (AFM) (Fig. S13), supplemented by previously published data.[10]

**Nano Device Fabrication:** Gold-titanium (Au-Ti) alloy terminals were deposited onto the atmospheric-treated films, leaving a 25 μm gap between terminals for inductance, capacitance, and resistance (LCR) measurements by the 2-probe method (Fig. S14, S15).

**Alloy thin film deposition on PI:** Finally, Pt-Ce alloy thin films were deposited onto polyimide (PI) substrates using the same sputtering process, followed by atmospheric treatments under the optimal condition at 300 °C for 30 minutes (see FE-SEM images in Fig. S16-S18). Au-Ti terminals were then deposited to the film with a 1 mm gap between terminals using the electron-beam evaporator (MB-501010). The electrical resistance of the terminated film was quantified by the 4-probe method (Fig. S19), subjected to repeated bending up to 1000 cycles at diameters as small as 1.5 mm.

## ASSOCIATED CONTENT

Supporting Information

## CONFLICTS OF INTEREST

There are no conflicts to declare

## ACKNOWLEDGMENTS

This work was supported by "Advanced Research Infrastructure for Materials and Nanotechnology in Japan (ARIM)" of the Ministry of Education, Culture, Sports, Science and Technology (MEXT)—proposal number JPMXP1223NM5363.